\documentclass[10pt]{article}
\usepackage{amsfonts}
\newcommand{\sect}[1]{\setcounter{equation}{0}\section*{#1}}
\renewcommand{\theequation}{\arabic{section}.\arabic{equation}}
\def\bea{\begin{eqnarray}}
\def\eea{\end{eqnarray}}

\def\beann{\begin{eqnarray*}}
\def\eeann{\end{eqnarray*}}
\def\beq{\begin{equation}}
\def\eeq{\end{equation}}
\def\ba{\begin{array}}
\def\ea{\end{array}}
\def\ben{\begin{enumerate}}
\def\een{\end{enumerate}}
\def\f{\phi}
\def\m{\mu}
\def\pd{\partial_}
\def\pu{\partial^}
\def\p{\psi}

\def\n{\nu}

\def\d{\delta}

\def\st{{}^\star}

\def\nn{\nonumber \\}

\renewcommand{\theequation}{\thesection.\arabic{equation}}

\font\mybb=msbm10 at 11pt

\def\bb#1{\hbox{\mybb#1}}

\def\bR {\bb{R}}

\def\e  {\epsilon}

\def\cL{{\cal L}}

\date{}

\begin{document}

\thispagestyle{empty}
\begin{center}
\vspace {.7cm} {\large {\bf{Brane Solitons for $G_2$ Structures in
Eleven-Dimensional Supergravity Re-Visited}} }
 \vskip 1.0truecm
 \vskip 1.0truecm
  {\large{\bf{
Jan Gutowski${}^1$ and George  Papadopoulos${}^2$}}}
 \vskip 2.0truecm
 {\normalsize{\sl 1. Department of Physics}}
\\ {\normalsize{\sl Queen Mary University of London}}
\\ {\normalsize{\sl  Mile End, London E1 4NS.}}
\vskip 1.0truecm {\normalsize{\sl 2. Department of Mathematics}}
\\ {\normalsize{\sl King's College London}}
\\ {\normalsize{\sl Strand}}
\\ {\normalsize{\sl London WC2R 2LS}}

\vskip 2.0truecm
{\bf {Abstract}}
\end{center}

We investigate the four-dimensional supergravity theory obtained from the
compactification of eleven-dimensional supergravity on a smooth
manifold of $G_2$
holonomy. We give a new derivation for the K{\"a}hler potential associated
with the scalar kinetic term of the $N=1$ four-dimensional theory.
We then examine some solutions of the four-dimensional theory which
 arise from wrapped M-branes.

\section{Introduction}

Considerable interest has been shown recently in compactifications of
M-theory on manifolds which have holonomy group $G_2$.
This is primarily because the compactified theory exhibits a realistic
amount of supersymmetry in four dimensions {\cite{paptown}}. However,
it has been known for some time that for compactifications of
eleven-dimensional supergravity on smooth   $G_2$ manifolds
the four-dimensional effective theory does not have charged particles.
Recently charged particles have been introduced
by considering compactifications on singular $G_2$ manifolds
{\cite{dualc}}, {\cite{cvt}}, {\cite{duald}}. This has led to renewed interest in
the properties of $G_2$ compactifications.
This has also motivated the investigation
of $G_2$ singularities and has led to the construction
of several new non-compact $G_2$ manifolds {\cite{andreas}},
and {\cite{gary}} and references within.

In this paper we shall re-derive the K\"ahler potential of
the N=1 four-dimensional supergravity action obtained from
the compactification of eleven-dimensional supergravity on a smooth $G_2$
manifold. Geometrically this gives a K\"ahler metric on the tangent
space of the moduli space of $G_2$ structures.
This K\"ahler potential has been obtained previously in {\cite{paptown}},
{\cite{harvey}}, {\cite{papa}} and {\cite{beasl}}.
 This was done by computing first the
terms quadratic in the 4-form field strength obtained
 on making a Kaluza-Klein
ansatz. Then $N=1$ supersymmetry together with the
results of {\cite{hitchin}} were used to determine the K\"ahler metric.
In this paper we explicitly derive an expression for
the K\"ahler potential of
the four-dimensional supergravity action from the eleven-dimensional
Einstein-Hilbert term, using some results of {\cite{joyce}}.
This new derivation is similar in spirit to the calculation
of the moduli metric of supersymmetric multi-black hole solutions;
see for example {\cite{shir}}, {\cite{michelson}}, {\cite{gutpapc}}. One
of the advantages of this new approach is that the
K\"ahler structure on the tangent bundle of the moduli space of
$G_2$ structures is obtained directly without making an explicit
use of supersymmetry in four dimensions. Our computation will be used
to compare a normalization factor in the definition of the
K\"ahler potential.

The plan of this paper is as follows. In section two we present
the computation of the K\"ahler potential of the
four-dimensional theory obtained on compactification by
making use of various symmetries of the $G_2$ manifold. In
section three we examine some black hole and domain wall
solutions of the resulting four dimensional
effective theory. Some useful $G_2$ identities are presented in
the Appendix.

\section{Compactifications of Eleven-Dimensional Supergravity}

The bosonic part of the eleven-dimensional supergravity Lagrangian is
\beq
\cL = \sqrt{h} R -{1 \over 2}F \land {}^\star F +{1 \over 6}C \land F
\land F \ ,
\label{sugraact}
\eeq
where  $h$ is the eleven dimensional spacetime
metric, $R$ is the eleven
dimensional Ricci scalar, $C$ is a 3-form gauge potential with associated
field strength $F=dC$. The compactification ansatz that we
shall consider is
\beq
ds^2 = ds^2 (\bR^{1,3}) + ds^2_{N} \ ,
\label{unpet}
\eeq
with $F=0$,
where $N$ is a compact smooth 7-manifold with $G_2$ holonomy. In particular,
$N$ is Ricci-flat,
and there exists a covariantly constant 3-form $\f$ on $N$. Some algebraic
properties  of this 3-form are presented in the Appendix.
The dimension of the moduli space of $G_2$ structures of the
manifold $N$ is $b_3(N)$.
The solution ({\ref{unpet}})  preserves $1 \over 8$ of the 32
 supersymmetries of eleven-dimensional supergravity.
Hence, we expect that the low energy four-dimensional theory is
 a  $N=1$, four-dimensional supergravity.

In order to probe the low energy dynamics of this solution,
we shall allow the moduli
$s^i$ to depend on
the spacetime co-ordinates $x^M$ as $s^i \to s^i (x^M)$.
Specifically, we shall take the perturbed spacetime metric to
be
\beq
ds^2 = g_{MN} dx^M dx^N + g_{ab} dy^a dy^b \ ,
\label{petmet}
\eeq
where we have split the spacetime co-ordinates in a 4+7 manner according to
$x^\m = \{ x^M \ , \ y^a \}$ for $M,N =0, \dots , 3$ and
$a, b=1, \dots , 7$; and $g_{ab}=g_{ab}(y,s^i(x^M))$, $g_{MN}=g_{MN}(x^L)$.
Here $g_{ab}$ depends on $x^M$ only via the moduli $s^i$, so that
at fixed $x^M$, $g_{ab} dy^a dy^b$ is a metric with $G_2$ holonomy.
We shall assume that all components of the metric vary slowly with regard
to $x^M$, and we shall refer to quantities being of first or second
order with respect to derivatives of $x^M$.

To proceed, it is necessary to solve the field
equations up to first order, and to determine
the effective action up to second order. Solving the field equations up
to first order ensures that the back reaction of the geometry
to the varying moduli is taken into account. In what follows,
we shall concentrate on the Einstein equations.
We remark that perturbing the moduli in the 3-form potential $C$
produces a first order 4-form field strength, and hence, in order to
solve the first order Einstein equations, it suffices to solve the
Ricci-flatness condition
\beq
R_{\m \n}=0
\label{rici}
\eeq
up to first order. The only first order terms contributing
to the Ricci tensor arise in the $R_{aM}$ components. This implies that
\beq
\nabla_d W_M{}^d{}_a - \nabla_a W_M{}^d{}_d =0
\label{covpet}
\eeq
where $\nabla_a$ is the covariant derivative with respect to the
 (perturbed) metric $g_{ab}$ on $N$ and
\beq
W_{Mab} = \cL_{\partial \over \partial x^M} g_{ab}\ ;
\label{simpl}
\eeq
$\cL$ is the Lie derivative.

To investigate the consequences of  ({\ref{covpet}}), it is convenient to
express $W_{Mab}$ in terms of the 3-forms
$\p_M = \cL_{\partial \over \partial x^M} \f$
and $\chi_M= \nabla_M \f$, where here $\nabla_M$ denotes
the covariant derivative with respect to the (perturbed)
eleven-dimensional metric.
Note that up to first order, the only non-vanishing components of
$\p_M$ and $\chi_M$ are $\p_{Mabc}$ and $\chi_{Mabc}$.
Next, taking the Lie derivative of  (\ref{metb}) in the
 appendix, we find that
\beq
W_{Mab} = {1 \over 6} \p_{Macd} \f_b{}^{cd} +
{1 \over 6} \f_a{}^{cd} \p_{Mbcd}
 +{1 \over 3} \pd{M} g^{cf}
 \f_{acd} \f_{bf}{}^d \ .
\label{diffy}
\eeq
For use later observe that contracting (\ref{diffy})
with $g^{ab}$, one obtains
\beq
\pd{M} \log \det g_7 = {1 \over 9} \p_{Mdef} \f^{def} \ .
\label{detvar}
\eeq
Furthermore, using (\ref{isitanygood}), the last term in (\ref{diffy})
can be rewritten as
\beq
\pd{M} g^{cf} \f_{acd} \f_{bf}{}^d = W_{Mab} - g_{ab}
\pd{M} \log \det g_7 \ .
\label{reasb}
\eeq
Substituting this expression back into (\ref{diffy}), one obtains
\beq
W_{Mab} = {1 \over 4} \big( \p_{Macd} \f_b{}^{cd} +
 \f_a{}^{cd} \p_{Mbcd} \big)
-{1 \over 2} g_{ab} \pd{M} \log \det g_7 \ .
\label{diffyb}
\eeq
It follows that
\beq
\nabla^b W_{Mab} - \nabla_a W_M{}^b{}_b = {1 \over 4}
 \nabla^b \big( \f_a{}^{cd}
\p_{Mbcd} - \p_{Macd} \f_b{}^{cd} \big) \ .
\label{final}
\eeq
Using the relation
\beq
\chi_{Mabc} ={1 \over 4} \p_{Mabc}-{1 \over 24}
 \f_{abc} \p_{Mdeh} \f^{deh}
-{3 \over 8} \p_{M de[a} \st \f_{bc]}{}^{de}
\label{covdif}
\eeq
between $\chi_{Mabc}$ and $\p_{Mabc}$, the condition  (\ref{covpet})
can be written as
\beq
\st \f \land \d \chi_M =0 \ ,
\label{formident}
\eeq
where $\d$ is the adjoint of $d$ defined on $\Lambda^* (N)$.

To continue, recall that $\Lambda^2(\bR^7)$  decomposes into two
irreducible $G_2$ representations, $\Lambda^2(\bR^7) = \Lambda^2{}_7 \oplus
\Lambda^2{}_{14}$. The seven-dimensional
 representation $\Lambda^2{}_7$ arises  by
contracting a two-form with the parallel three-form $\phi$ to yield
a vector.
Then the condition  (\ref{formident}) implies that $ \pi_7 \d \chi_M =0$,
where $\pi_7$ is the projection onto $ \Lambda^2{}_7$.

Now we shall show that the field equations  (\ref{covpet}) are solved if
$\p_M$ is harmonic. For this we choose the gauge fixing condition
$\pi_7 \d \p_M=0$. This is a good gauge fixing condition because it
can be shown
that it is equivalent to requiring that
$\cL_{\partial \over \partial x^M} \f$
is $L^2$-orthogonal to $\cL_{\partial \over \partial y^a} \f$, ie
 the deformations
along the moduli directions are orthogonal to the deformations induced
by the diffeomorphisms of $N$. It turns out that the closure of $\p_M$
together with the condition $\pi_7\d\p_M=0$ imply that $\p_M$ is harmonic.
In particular $\p_M$ is coclosed and so the first term
in the right-hand-side
of (\ref{final}) vanishes.  Using the fact that
$\p_M$ is harmonic, it can be shown that
  $\p_{Mabc} \f^{abc}$ is independent of $y^a$ and so the second term
  in the right-hand-side of (\ref{final}) vanishes as well.

It remains to compute the second order effective action from
(\ref{sugraact}) .
We shall concentrate on the Einstein-Hilbert part of the action, as
this is sufficient to determine the K\"ahler
 potential of the four-dimensional $N=1$ theory.
The zeroth and first order parts in four-dimensional
spacetime derivatives of the eleven-dimensional action vanish.
The second order part
simplifies to
\bea
S_{EH}=\int d^{11}x \sqrt{-\det {g_4}} \sqrt{\det{g_7}}
\big[R^{(4)} -{1 \over 4}g^{MN}\big(W_M{}^{ab}W_{Nab}-
W_M{}^a{}_aW_N{}^b{}_b\big)\big]\ .
\label{seccv}
\eea

Using the expression for $W_{Mab}$ in terms of $\psi_M$ in
(\ref{diffyb}),  we find that the
 eleven-dimensional Einstein-Hilbert
term of the action can be written as
\bea
S_{EH} &=& \int d^{11}x \sqrt{-\det {g_4}} \sqrt{\det{g_7}}
\big[R^{(4)}
-{1 \over 16} g^{MN} \p_{Mabc}\p_N{}^{abc}
\nn
&-&{1 \over 32} g^{MN} \st \f^{cdeh} \p_{Macd} \p_N{}^a{}_{eh} +
 {1 \over 864} g^{MN}
\big(\f^{abc} \p_{Mabc} \big) \big(\f^{deh} \p_{Ndeh} \big) \big]\ .
\label{simpehb}
\eea

This expression simplifies even further when we recall that the 3-forms
on $N$ decompose into three $G_2$ irreducibles:
$\Lambda^3(N) = \Lambda^3{}_1 \oplus
\Lambda^3{}_{7} \oplus \Lambda^3{}_{27}$
and the harmonic 3-forms decompose as
$H^3 (N, \bR ) = H^3{}_1 (N, \bR ) \oplus H^3{}_{27} (N, \bR)$.
From this it follows that $\pi_7 \p_M=0$, which implies that
$\f \wedge \p_M=0$. Hence we obtain

\bea
S_{EH} &=& \int d^{11}x \sqrt{-\det {g_4}} \sqrt{\det{g_7}}
\big[R^{(4)}
-{1 \over 12} g^{MN} \p_{Mabc}\p_N{}^{abc}
\nn
& + &  {1 \over 216} g^{MN} \big(\f^{abc} \p_{Mabc} \big)
\big(\f^{deh} \p_{Ndeh}
 \big) \big]\ .
\label{simpehbii}
\eea

To continue, it is convenient
to define $\Theta = \int d^7 y \sqrt{\det g_7}$, so that
\beq
\p_{Mabc} \f^{abc} = 18 \Theta^{-1} \pd{M} \Theta \ .
\label{cconfr}
\eeq
To put the four-dimensional action into the Einstein frame,
we make the conformal re-scaling
\beq
{\hat{g}}_{MN}= \Theta g_{MN} \
\label{confresc}
\eeq
of the
four-dimensional
metric.
Then the Einstein-Hilbert action is written as

\beq S_{EH} = \int d^{4}x \sqrt{-\det {g_4}} \big[R^{(4)} -
{1 \over 2} \Theta^{-1} g^{MN} (\p_M , \p_N) \big]\ ,
\label{simpehbiii} \eeq where we have dropped the $\ \hat{} \ $
and \beq (\tau , \kappa) \equiv {1 \over 6} \int d^7 y \sqrt{ \det
g_7 } \ \tau_{abc} \kappa^{abc} \label{innerprod} \eeq
for 3-forms $\tau, \kappa \in \Lambda^3(N)$.
 After a
calculation similar to the one presented in \cite{papa} and
\cite{beasl}, we obtain the K\"ahler potential \beq K= -3 \log
\Theta \ , \label{kpot} \eeq
of the sigma model matter in the
effective $N=1$, four-dimensional supergravity theory. In fact
(\ref{kpot}) should be thought of as the K\"ahler potential on the
tangent bundle of the  moduli space of $G_2$ structures on the
compact manifold $N$. Note that the normalization factor of the
K\"ahler potential is that of \cite{beasl} and it differs from
that of \cite{papa} by a factor of seven.

The remaining couplings of the  four-dimensional effective supergravity
have been described extensively in the literature  {\cite{paptown}}.
The bosonic fields are $b_3(N)$ complex scalars
$z^i={1 \over 2}(s^i-ip^i)$, $i=1, \dots , b_3(N)$ and $b_2(N)$
Maxwell fields $F^a = d A^a$, $a=1, \dots , b_2(N)$, where the
real scalars $p^i$ and the Maxwell fields arise from the Kaluza-Klein
reduction of the three-form gauge potential of eleven-dimensional
supergravity given by
\beq
C= p^i \phi_i + A^a \wedge \omega_a
\label{kkreduc}
\eeq
where $\{ \phi_i \}$ is a basis for $H^3(N, \bR)$ and $\{ \omega_a \}$
is a basis for $H^2(N, \bR)$. The total effective action obtained
via this reduction combined with the calculation given above is then
embedded within the $N=1$ four-dimensional supergravity as
\beq
S_{G_2} = \int \ d^4 x \sqrt{-g} \big[R-2 \gamma_{i \bar{j}} \pd{M} z^i
\pu{M} {\bar{z}}^j -{1 \over 2} {\rm Re} h_{ab} F^a{}_{MN} F^{b MN}
+{1 \over 2} {\rm Im} h_{ab} F^a{}_{MN} {}^\star F^{b MN} \big]
\label{totalactb}
\eeq
where $ \gamma_{i \bar{j}}= {\partial^2 K \over \partial z^i
\partial {\bar{z}}^j}$ and $h_{ab} = h_{ab}(z^i)$ is holomorphic
in $z^i$.
In particular, from the Kaluza-Klein reduction of the
term $-{1 \over 2} F \wedge \star F$ in the eleven-dimensional
supergravity action we obtain the following four-dimensional terms
\beq
\int \ d^4 x \sqrt{-g} \big[-{1 \over 2} \Theta^{-1} g^{MN} \pd{M}p^i
 \pd{N} p^j  (\phi_i , \phi_j)
-{1 \over 4} s^i C_{iab} F^a{}_{MN} F^{b MN} \big]
\label{kka}
\eeq
where $C_{iab}$ are
 real topological intersection numbers given by
\beq
C_{iab}= \int_N \phi_i \wedge \omega_a \wedge \omega_b \ ,
\label{topol}
\eeq
and we have made use of the identity
$\omega_a \wedge {}^\star \omega_b = \omega_a \wedge \omega_b \wedge \f$
which follows from ({\ref{projsev}}). From the eleven dimensional term
${1 \over 6} F \wedge F \wedge C$
we obtain the four-dimensional term
\beq
\int \ d^4 x \sqrt{-g} \big[
-{1 \over 4} p^i C_{iab} F^a{}_{MN} {}^\star F^{b MN} \big]\ .
\label{kkb}
\eeq
Hence, comparing ({\ref{kka}}) and ({\ref{kkb}}) with ({\ref{totalactb}})
 we find that
${\rm Re} h_{ab} = {1 \over 2} s^i C_{iab}$ and
 ${\rm Im} h_{ab}=-{1 \over 2} p^i C_{iab}$;
and so $h_{ab}=z^i C_{iab}$ where $z^i={1 \over 2}(s^i-ip^i)$.

We remark that a potential term of the form {\cite{wess}}
\beq
S_{pot} = - \int \ d^4 x \sqrt{-g}\big( e^K \big[ \gamma^{i \bar{j}}D_i f
 D_{\bar{j}} {\bar{f}}-3 |f|^2 \big] +{1 \over 2} D_a D^a \big)
\label{potterm}
\eeq
may in principle be added to the $G_2$ action given
in ({\ref{totalactb}}), where
$f=f(z^i)$ is a holomorphic superpotential, $D_i f = f K_i +
{\partial f \over \partial z^i}$,
 $K_i = {\partial K \over \partial z^i}$; $D_a$ is a
 Fayet-Iliopoulos term and $D^a = {\rm Re} h^{ab}D_b$
 where $ {\rm Re} h^{ab}$
is the inverse of $ {\rm Re} h_{ab}$. The action given by $S_{G_2}+S_{pot}$
possesses $N=1$ supersymmetry in four dimensions; however, it is not
generically possible to obtain the potential term given above through the
$G_2$ compactification described here.

\section{Solutions of the Effective Theory}

\subsection{Dilatonic Black Holes}

The simplest possible black-hole type of solution
with active scalars is that for which the only non-vanishing modulus
field is the volume of the compact space. The ansatz for such
a black hole have been given in \cite{papa}. In particular one takes
$ds^2=-A^2 dt^2+B^2 (dr^2+r^2 ds^2(S^2))$,   $s^i = s c^i$ with $c^i$
constant; $p^i=0$
and $A^a= C^a dt$. The potential term is set to vanish.
The solution can be found by a direct substitution into the field
equations of the truncated action
\beq
S'_{G_2} = \int \ d^4 x \sqrt{-g} \big[R-2 \gamma_{i \bar{j}} \pd{M} z^i
\pu{M} {\bar{z}}^j -{1 \over 2} {\rm Re} h_{ab} F^a{}_{MN} F^{b MN}\big]\ ,
\label{totalactc}
\eeq
where the $p F \wedge F$ terms have been truncated.
The spacetime metric, gauge field strengths and scalars are given by
\bea
ds^2&=&-L^{{7 \over 4}} F dt^2+  L^{-{7 \over 4}}
 [F^{-1}dr^2+ r^2 ds^2(S^2)]
\nn
F^a &=& H^a dL \wedge dt
\nn
s&=& \rho L^{-{1 \over 4}}
\label{nonexta}
\eea
where $H^a$ and $\rho$ are constants, and
\bea
L^{-1}&=&1+{4\kappa^2 \mu\over r}
\nn
F&=&1-{4\mu\over r}\ ,
\label{nonextb}
\eea
for constant $\mu$, $\kappa$.  It is also convenient to define the constants
${\cal{H}}_{ab}=c^i C_{iab}$, and set $K_i=s^{-1} \lambda_i$
where $\lambda_i$ are constants such that $c^i \lambda_i=-7$,
 as a consequence of the
homogeneity properties of the volume of the $G_2$ manifold.
 It is straightforward to
show that in order for the scalar equations to be satisfied,
 the $\lambda_\ell$ must be constrained as
\beq
C_{\ell a b}H^a H^b = -{(\kappa^2+1) \over 2 \rho \kappa^2} \lambda_\ell \ ,
\label{topoconstr}
\eeq
and given this constraint, the Einstein equations are also satisfied.
Note that $\mu$ is the non-extremality parameter of the
black hole. Taking the limit $\mu\rightarrow 0$ with $4\kappa^2\mu=Q$
finite, the resulting solution is an extremal black hole. In fact it is one
of the dilatonic black holes found in  {\cite{shir}}.
 To see this, observe that a consistent truncation of the four-dimensional
  action ({\ref{totalactc}}) associated
with this ansatz, after an appropriate rescaling  of the Maxwell field, is
\beq
S'_{G_2}= \int d^{4}x \sqrt{-\det {g_4}}
\big[R^{(4)}
-{7 \over 2} s^{-2} |\nabla s|^2 -s F^2 \big]\ ,
\label{simpehbiiivb}
\eeq
where $F=dA$ is the 2-form Maxwell field strength.
 This action is equivalent to the
Einstein-Maxwell action for dilatonic black holes given in {\cite{shir}}
\beq
S = \int d^{4}x \sqrt{-\det {g_4}}
\big[R^{(4)}-2 |\nabla \sigma|^2 -e^{-2a \sigma} F^2 \big]\ ,
\label{emax}
\eeq
on setting $s=e^{-{2 \over \sqrt{7}} \sigma}$
together with $a= {1 \over \sqrt{7}}$.
 The multi-black hole extreme solutions of (\ref{simpehbiiivb}) are
given as
\beq
ds_4{}^2 = H^{-{7 \over 4}} dt^2+ H^{7 \over 4}d{\bf{x}}^2
\label{bhmet}
\eeq
together with $s=H^{1 \over 4}$ and $A= {\sqrt{7 \over 8}}H^{-1} dt$
where
\beq
H=1+ \sum_{A=1}^N {\mu_A \over |{\bf{x}}-{\bf{z}}_A|}
\label{harmfunct}
\eeq
is a harmonic function\footnote{The power of the
harmonic function in these solutions differs from that given in \cite{papa}
because of the change of normalization of the K\"ahler potential.}.
Note that the volume of the $G_2$ manifold blows
up at the poles of $H$, which are
curvature singularities of the four-dimensional
spacetime solution.

There are also magnetic black holes associated with $G_2$ compactifications
and their relation to wrapped branes has been described in \cite{papa}.
The metric for these black holes is the same as that of the electric
black holes above, but the modulus
 scalar for the magnetic black holes is the
inverse of the electric black hole modulus scalar.
Both the electrically and magnetically charged black holes break all of
the $N=1$ four-dimensional supersymmetry.

\subsection{Supersymmetric Domain Walls}

In order to find domain wall solutions preserving
half of the $N=1$ supersymmetry, it is necessary to introduce
a potential term into the action {\cite{supdw}}.
Although such a term cannot be obtained from
the classical compactification described here, one
way of obtaining this potential is to
add additional fluxes to the $G_2$ manifold {\cite{guk}}.
However, it has been
argued that this is inconsistent with the assumption that the $G_2$
manifold is compact {\cite{beck}}.
Alternatively, one may include quantum
corrections which arise from the wrapping
 of Euclidean M2-brane instantons on
associative cycles {\cite{harvey}}, {\cite{misr}}. A superpotential for this
 configuration has been conjectured in {\cite{oog}} which is of the form
\beq
f = \m \sum_{k=1}^{\infty} {1 \over k^2} e^{k\lambda_i z^i}
\label{insstantpot}
\eeq
for constants $\m$ and $\lambda_i$.

To describe domain wall type solutions we set the gauge fields and
the Fayet-Iliopoulos terms to vanish, and take the spacetime
metric to be \beq ds^2 = B(y)^2 \big[ds^2(\bR^{1,2}) +dy^2 \big]
\label{domnwallmet} \eeq with $z^i = z^i (y)$. The generic
constraint equations for such a solution which preserves half of
the supersymmetry have been given in in {\cite{gutt}}; they are
\bea \pd{y}B &=&-e^{i \phi} B^2 e^{K \over 2} {\bar{f}}
\nn
\pd{y} z^i &=& -B^{-1} \pd{y}B {\bar{f}}^{-1}
\gamma^{i \bar{j}} D_{\bar{j}} {\bar{f}}
\nn
\pd{y} \big[2 \Psi +iK \big] &=& 2i K_i \pd{y} z^i\ ,
\label{susycons}
\eea
where $\Psi$ is an angular variable obtained from solving the Killing
spinor equations. To solve these equations we shall set $p^i=0$.
The third
equation in ({\ref{susycons}}) implies that $\pd{y} \Psi=0$.
Hence it is convenient to rescale $f = e^{i \Psi_0}F$ where
$\Psi = \Psi_0 = {\rm const}$. Then ({\ref{susycons}}) simplifies to
\bea
\pd{y} B &=&-B^2 e^{K \over 2} {\bar{F}}
\nn
\pd{y} z^i &=& Be^{K \over 2} \gamma^{i \bar{j}} D_{\bar{j}} {\bar{F}}
\label{susysimp}
\eea
To proceed we set $s^i = s(y)c^i$ for constant $c^i$.
 As a consequence of
the homogeneity of the $G_2$ volume, we also have $K_i= \lambda_i s^{-1}$
for constant $\lambda_i$, where $c^i \lambda_i =-7$. Furthermore,
$\Theta = \mu s^{7 \over 3}$ for constant $\mu>0$. If we set
$F= F(-{2 \over 7} z^i \lambda_i)$, so that $F|_{p^i=0,s^i=s c^i}=P(s)$
then ({\ref{susysimp}}) simplifies to
\bea
\pd{y} B^{-1} &=& \mu^{-{3 \over 2}} s^{-{7 \over 2}} P
\nn
\pd{y} s &=& {4 \over 7} \mu^{-{3 \over 2}}s^2 (s^{-{7 \over 2}}P)'
\label{susysimpb}
\eea
where $'= {d \over ds}$. Hence, defining $W=s^{-{7 \over 2}}P$, we
note that we can write $B$ in terms of $s$ as
\beq
B= e^{-{7 \over 4} \int {W \over s^2 W'} ds}
\label{bsol}
\eeq
and given such a solution, the behaviour of $s$ is determined by the
equation
\beq
\pd{y} B^{-1} = \mu^{-{3 \over 2}}W\ .
\label{finaldw}
\eeq
We note that just as for the case of the black holes, the power of
the exponential in ({\ref{bsol}}) is altered due to the change
of normalization of the K\"ahler potential.

We have proposed a  new method to compute the K\"ahler potential of scalars
of four-dimensional effective theory associated with
 D=11 supergravity compactifications on compact smooth $G_2$ manifolds.
This method may be proved useful to compute the K\"ahler potential
of the effective theory when the underlying  $G_2$ manifold in not compact.

\vskip 1cm \noindent{\bf Acknowledgements:}  J. G. is supported by an
 EPSRC postdoctoral grant.
 G.P. is
supported by a University
Research Fellowship
from the Royal Society.
 This work is
partially supported by SPG grant PPA/G/S/1998/00613.

\appendix

\makeatletter
\renewcommand{\theequation}{a.\arabic{equation}}
\@addtoreset{equation}{section} \makeatother
\sect{Appendix : Some Useful $G_2$ Identities}

If $N$ is an integrable $G_2$ manifold, then there exists
a covariantly constant
3-form $\f$,
 which, in an appropriate
 orthonormal frame $\{ e^a :a=1, \dots , 7 \}$ may be written as
\bea
\f  &=&e^1 \land e^2 \land e^3 + e^1 \land \big(e^4 \land e^5
- e^6 \land e^7 \big)
+ e^2 \land \big(e^4 \land e^6
+ e^5 \land e^7 \big)
\nn
&+&e^3 \land \big(e^4 \land e^7 - e^5 \land e^6 \big) \ ,
\label{threeform}
\eea
and the dual 4-form is
\bea
{}^\star \f &=& e^4 \land e^5 \land e^6 \land e^7+
e^2 \land e^3 \land \big(e^6 \land e^7 - e^4 \land e^5 \big) +
e^1 \land e^3 \land \big(e^4 \land e^6 + e^5 \land e^7 \big)
\nn
& + & e^1 \land e^2 \land \big(e^5 \land e^6 - e^4 \land e^7 \big) \ ,
\label{dualform}
\eea
where $\st$ is the Hodge dual defined with respect to $g_7$ and we
fix $\e^{1234567}=+1$.
The $G_2$ metric is obtained from the 3-form $\f$ via
the non-linear relation
\beq
g_{ab} = - {1 \over 24} \f_{acd}\f_{bef}{}^\star \f^{cdef}\ .
\label{hitch}
\eeq
In addition, we have
\beq
g_{ab} = {1 \over 6} \f_{acd} \f_b{}^{cd}\ .
\label{metb}
\eeq
Another useful identity is
\beq
\f_{a c d} {}^\star \f^a{}_{b h f}  =- \big[g_{c b} \f_{d h f}+g_{c h}
\f_{d f b} + g_{c f}\f_{d b h}
 - g_{d b} \f_{c h f}-g_{d h}
\f_{c f b} - g_{d f}\f_{c b h} \big]\ ,
\label{identi}
\eeq
which implies
\beq
\f_{acd} {}^\star \f^{ac}{}_{hf} = -4 \f_{dhf}\ .
\label{identib}
\eeq
We also have the identity
\beq
\f_{dab} \f^d{}_{cf} = g_{ac} g_{bf}-g_{af} g_{bc} -\st \f_{abcf}\ ,
\label{isitanygood}
\eeq
and for any 3-form $\theta$,
\beq
-{1 \over 3} \f_{abc} \theta_{deh} \f^{deh}-
2 \theta_{de [a} \st \f_{bc]}{}^{de}
+ \theta_{deh} \f^{eh}{}_{[a} \f_{bc]}{}^d =0 \ .
\label{thrfrmident}
\eeq
Note also that if $\omega \in \Lambda^2(N)$ is a 2-form with
$\pi_7 \omega=0$ then the constraint
\beq
\f_a{}^{bc} \omega_{bc}=0 \ ,
\label{projseva}
\eeq
which is satisfied if $\omega$ is harmonic, is equivalent to
\beq
\omega_{ab} = {1 \over 2} {}^\star \f_{ab}{}^{cd} \omega_{cd}
\label{projsev}
\eeq
as a consequence of ({\ref{isitanygood}}).


\end{document}